\documentclass[10pt]{article}  
  
  \usepackage{makeidx}
  \usepackage{graphicx,graphics,color}
  \usepackage{latexsym,doc,fontenc,exscale}
 \usepackage{amsmath,amsfonts,amsthm,eucal,amssymb}

\begin{document}

 \title{ \textbf{Nonlinear dynamics of a position-dependent mass driven Duffing-type oscillator}}

\author{\textbf{Bijan Bagchi}\footnote{bbagchi123@rediffmail.com}, \textbf{Supratim
Das}\footnote{supratimiitkgp@gmail.com}, \textbf{Samiran Ghosh}\footnote{sran\_g@yahoo.com}
 and \textbf{Swarup Poria}\footnote{swarup\_p@yahoo.com}\\
\small{Department of Applied Mathematics, University of Calcutta},\\ \small 92 Acharya Prafulla Chandra Road, Kolkata-700009, India}
 \date{ }
\maketitle
 \begin{abstract}
 We examine some nontrivial consequences that emerge from interpreting a position-dependent mass (PDM) driven Duffing oscillator in the presence of a quartic
 potential. The propagation dynamics is studied numerically and sensitivity to the PDM-index is noted. Remarkable transitions from a limit  cycle to  chaos through period doubling and from a chaotic to a regular motion through intermediate periodic and chaotic  routes are  demonstrated.\\\\
   \noindent %
  \mbox{PACS number(s):  05.45.-a,  03.50.Kk,  03.65.-w}
  \end{abstract}
 \hspace{0.5cm} While position-dependent (effective) mass (PDM) quantum mechanical systems have repeatedly received attention in many areas of physics  (such as for instance, in the problems of compositionally graded crystals \cite{gel}, quantum dots \cite{ser}, nuclei \cite{rin},
 quantum liquids \cite{ari}, metal clusters \cite{pue} etc; see also \cite{bb1,bb2,mid,mus,cap,gang,ayd,chi} for theoretical developments and references therein), interest in classical problems having a PDM is relatively recent and rapidly developing subject \cite{car1,car2,car3,cru1,cru2,cru3,cru4,cru5,sg,lak1}. The simplest case of a PDM classical oscillator has been approached by several authors \cite{car3,cru1,cru2,cru3,cru4}. In what follows we shall demonstrate that a PDM ascribed Duffing oscillator provides an attractive possibility of defining a dynamical system that exhibits various features of bifurcations, chaos and regular motions. To be specific we will focus on the following equation of motion which emerges from a Lagrangian of the form \cite{lak2}
 \begin{equation}\label{lagrangian}
 L=\frac{1}{2}(\frac{1}{1+\xi x^{2}})(\dot{x}^{2}-\omega_{0}^{2}x^{2}),~~~\xi\in\mathbb{R}
 \end{equation}
 (where an overhead dot indicates a time derivative) namely,
 \begin{equation}\label{eq-motion-1}
 (1+\xi x^{2})\ddot{x}-\xi x\dot{x}^{2}+\omega_{0}^{2}x=0.
 \end{equation}
The above equation represents a special type of nonlinear oscillator which can be reduced to a first-order form by effecting a change in the variable $x$ \cite{lak2}. The latter equation can be immediately solved yielding periodic solutions. Hence, the absolute regularity of the motion does not depend on the specific value of $\xi$.

 Classically when the mass is position dependent, Newton's equation of motion gets modified to
 \begin{equation}\label{eq-motion-3}
 m(x)\ddot{x}+m'(x)\dot{x}^{2}=0
 \end{equation}
 (where the prime indicates a spatial derivative) in the absence of any external force term. When compared with (\ref{eq-motion-1})
 the following profile of the mass function comes out
 \begin{equation}\label{mass-functions}
 m(x):=\frac{1}{\sqrt{1+\xi x^{2}}}
 \end{equation}
 on ignoring the presence of the harmonic term $\omega_{0}^{2}x$ to effect such a comparison. In (\ref{mass-functions}) we have scaled the constant mass to unity. Notice that it is also possible to deal with other mass functions \cite{car2,cru4} but for concreteness we focus on  (\ref{mass-functions})
 in the rest of this paper.

 In \cite{cru4} an attempt was made to construct the underlying Lagrangian wherein the PDM system (\ref{eq-motion-3}) is acted upon by a force $F$ that could be dependent on position $x$, velocity $\dot{x}$ and time $t$ i.e.
 \begin{equation}\label{eq-motion-4}
 F(x,\dot{x},t):=\frac{dp}{dt}=m'(x)\dot{x}^{2}+m(x)\ddot{x}
 \end{equation}
 where $p=m(x)\dot{x}$ is the linear momentum. They have found that, in addition to  the kinetic energy $T$, a reacting thrust $\tilde{R}$ is at play as well
 \begin{equation}\label{KE}
 T:=\frac{1}{2}m(x)\dot{x}^{2},~~~~~~\tilde{R}(x,\dot{x},t):=-\frac{1}{2}m'(x)\dot{x}^{2}.
 \end{equation}
 where $\tilde{R}$ is essentially a non-inertial force. Assuming that $F$ can be split as $F=F(x)+\tilde{R}(x,\dot{x},t)$, where
 $F(x)$ is controlled by a scalar potential function $F=-\frac{\partial V}{\partial x}$, the Lagrangian form of (\ref{eq-motion-4}) obeys
 \begin{equation}\label{lag-eq}
 \frac{d}{dt}(\frac{\partial L}{\partial \dot{x}})-\frac{\partial L}{\partial x}=\tilde{R},~~~~L=T-V.
 \end{equation}
 Corresponding to the above $L$ which is $L=\frac{1}{2}m(x)\dot{x}^{2}-V(x)$, the Hamiltonian reads $H=\frac{p^{2}}{2m(x)}+V(x)$.
 These are of standard text-book forms\footnote{Quantum mechanically such forms are evidently non-Hermitian \cite{zno}} with $m=m(x)$.

 We stress that the non-potential force term $\tilde{R}$ has to depend on the velocities for otherwise they can be identified with the potential force $F(x)$. When written explicitly (\ref{lag-eq}) looks
 \begin{equation}\label{T-eq-1}
  \frac{d}{dt}(\frac{\partial T}{\partial \dot{x}})-\frac{\partial T}{\partial x}=-\frac{\partial V}{\partial x}+\tilde{R}.
 \end{equation}
 As an aside, the time rate of change of $T$ can be easily worked out as
 \begin{equation}\label{T-eq-2}
 \frac{dT}{dt}=\frac{d}{dt}(\frac{\partial T}{\partial \dot{x}}\dot{x})+(\frac{\partial V}{\partial x}-\tilde{R})\dot{x}+\frac{\partial T}{\partial t}
 \end{equation}
 from which, for a scleronomic system it at once follows that the time rate of energy $E(=T+V)$ is given by
 \begin{equation}\label{E-eq}
 \frac{dE}{dt}=\tilde{R}\dot{x}=-\frac{1}{2}m'\dot{x}^{3}
 =\frac{1}{2}\frac{\xi x\dot{x}^{3}}{(1+\xi x^{2})\sqrt{1+\xi x^{2}}}
  \end{equation}
 (\ref{E-eq}) speaks of the power of the non-potential force. This result is new for PDM systems and of interest. If the power is negative we encounter dissipative systems.

  The purpose of this communication is to develop a mathematical framework that enables us to study the evolution of the PDM system (\ref{eq-motion-4}) in the presence of an external periodic (non-autonomous) force
 with an additional damping term moving in a quartic potential. Towards this end we focus on an extended PDM equation of motion
 \begin{equation}\label{equation of motion}
 m(x)\ddot{x}+m'(x)\dot{x}^{2}+\omega_{0}^{2}x+\lambda x^{3}+\alpha \dot{x}=f\cos\omega t
 \end{equation}
 which reduces to (\ref{eq-motion-3}) in the absence of all the force terms.

  Some remarks are in order \cite{lak3,guc,chu}. For the constant mass case i.e. $m(x)=1$, (\ref{equation of motion}) goes over to a forced, damped Duffing oscillator which because of the presence of a double-well potential mimicks a magneto-elastic mechanical system.
 The latter is concerned with a beam placed vertically between two magnets with a fixed top end and free to swing at the bottom end. As soon as a velocity
 is enforced the beam begins to oscillate eventually coming  to rest at an equilibrium point. However the situation changes when a
 periodic force is applied :  stable fixed points or stable fixed angles no longer occur. In (\ref{equation of motion}) $\lambda$ is the governing
 parameter along with the periodic force $f\cos\omega t$ in the presence of a viscous drag of coupling strength $\alpha$.

Taking $m(x)$ as in (\ref{mass-functions}) we obtain from (\ref{equation of motion}) the coupled set of equations
\begin{eqnarray}\label{ds_1}
 \dot{x} &=& y  \nonumber\\
 \dot{y} &=& \frac{\xi x y^2}{1+\xi x^2} + \sqrt{1+\xi x^2}[ f\cos z -\omega_0^2 x -\lambda x^3 -\alpha y]\\
 \dot{z} &=& w \nonumber
 \end{eqnarray}
 \hspace{0.5cm}
 The remarkable behavior of the dynamical system (\ref{ds_1}) can be understood by examining the interplay between the amplitude $f$ of the periodic
 forcing term and the PDM parameter $\xi$. Various studies of the corresponding constant-mass case, i.e. when $\xi=0$, have been made in the
 literature (see specifically \cite{don}) and the nonlinear behavior demonstrated including the oscillation modes and the nonlinear resonances, both
 theoretically and experimentally. For the PDM case, we fix the parameter values to be $\omega=1.0,~\omega_{0}^{2}=0.25,~\alpha=0.2$ and $\lambda=1.0$ as is standard \cite{lak1}. Further, we will always assume $\xi\geq 0$. We plot in Figure 1 the phase diagrams for different values of $\xi$ including the constant-mass case of $\xi=0$. While in the
 latter situation we encounter limit-cycle oscillation, there is a drastic change in the dynamical behaviors as $\xi$ is increased. For instance, period-four oscillations are observed for $\xi=0.2$ which eventually give way to a chaotic behaviour both for $\xi=0.4$ and $\xi=0.6$ values.

 Bifurcation diagram of the system (\ref{ds_1}) for $x$ with respect to $f$ by  taking $\xi=0.5$ are presented in Figure 2. As soon as periodic force is applied, limit cycle oscillations gather in the range $0<f<2$. But with the increase of the amplitude of the forcing term, period-two oscillations set in and survive briefly up to $f=4$. Further stepping up of $f$ produces period-four and period-eight oscillations ultimately leading to a chaotic behaviour. But such a regime is short-lived because with the $f$-value going beyond $7.5$, a period-halving bifurcation appears yielding period-three
 oscillations for $f>8$.

  In Figure 3 we demonstrate the bifurcation sequence for $x$ but with respect to $\xi$ by taking $f=5.0$. As is evident, the limit cycle turns to period-four oscillations progressing to period-eight and so on running into a chaotic behaviour that lasts for a while before a regular motion takes over. The latter then again yields to a chaotic  dynamics and the entire motion subsequently  becomes a regular one around $\xi=1.9$. On the
  other hand, by carrying out a bifurcation analysis for $x$ with respect  to $\xi$ by taking $f=8.0$ we  find from Figure 4 that the constant-mass oscillator reveals a chaotic state. But we also notice that a subtle interplay between the parameters $f$ and $\xi$ produces complicated dynamics from a chaotic phase to periodic oscillations back again to a chaotic character and finally settling into a regular bahaviour.
\begin{figure}
\begin{center}
\includegraphics[width=1.\textwidth]{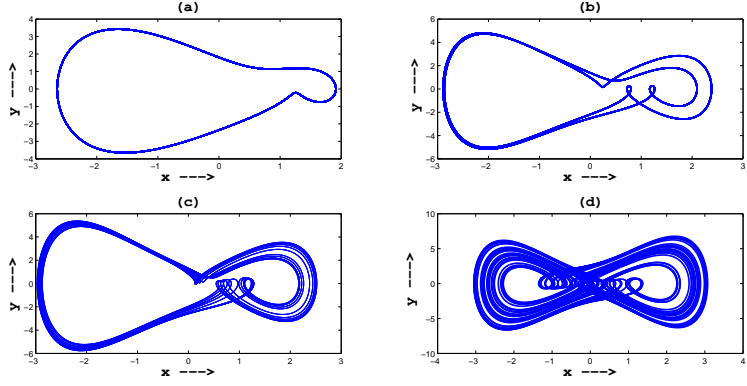}
\caption{ Phase diagram of equation (\ref{ds_1}) with $ f=5.0,~\omega=1.0,~\omega_0^2=0.25,~\alpha=0.2,~\lambda=1.0 $ for (a) constant mass $(\xi=0) $(b) PDM with $\xi=0.2$ (c) PDM with $ \xi=0.4 $ (d) PDM with $\xi=0.6$ }
\end{center}
\end{figure}

\begin{figure}
\begin{center}
\includegraphics[width=1.\textwidth]{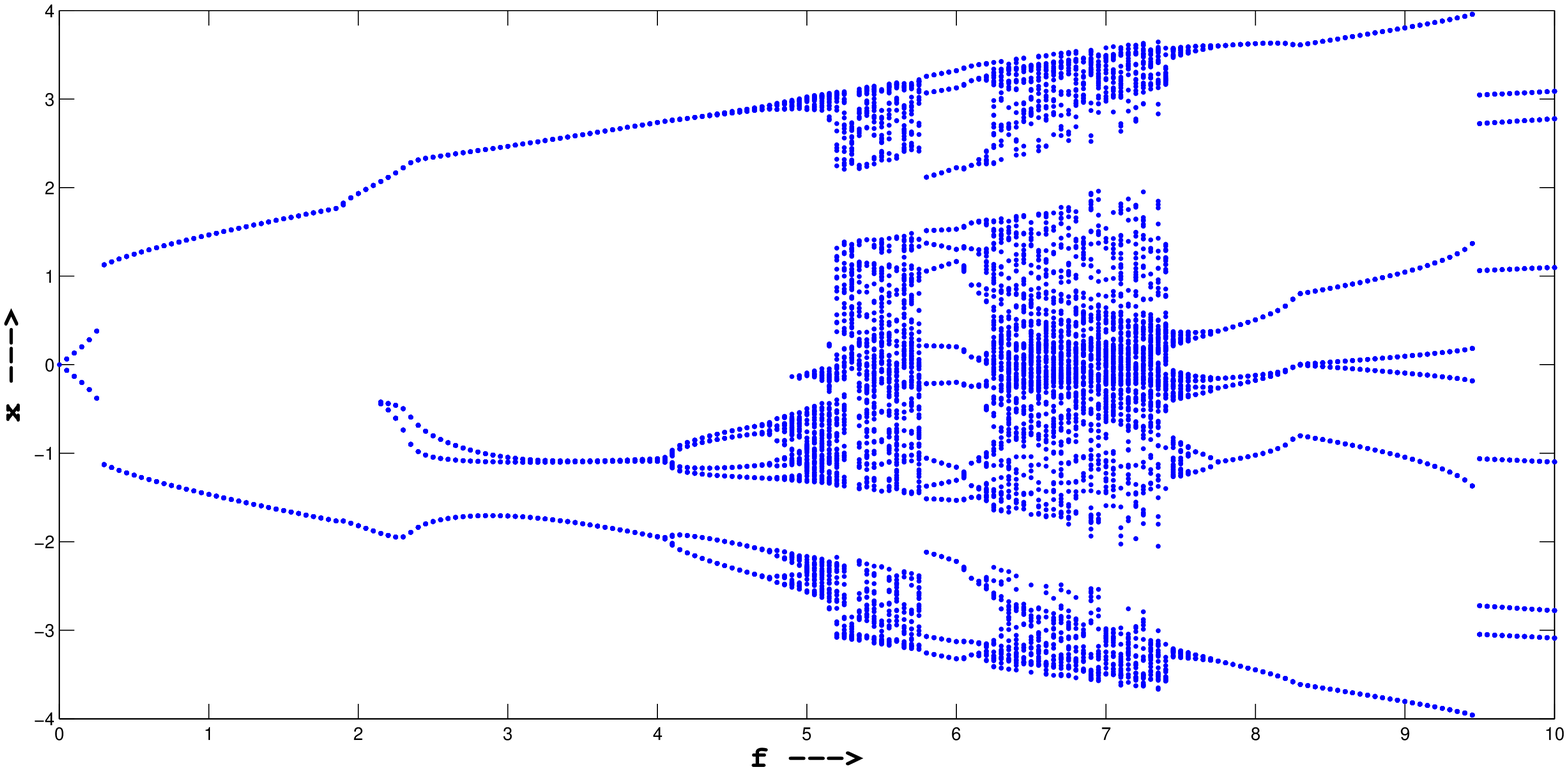}
\caption{ Bifurcation diagram of equation (\ref{ds_1}) with respect to $f$ for $\xi=0.5,~\omega=1.0,~\omega_0^2=0.25,~\alpha=0.2,~\lambda=1.0$ }
\end{center}
\end{figure}

\begin{figure}
\begin{center}
\includegraphics[width=1.\textwidth]{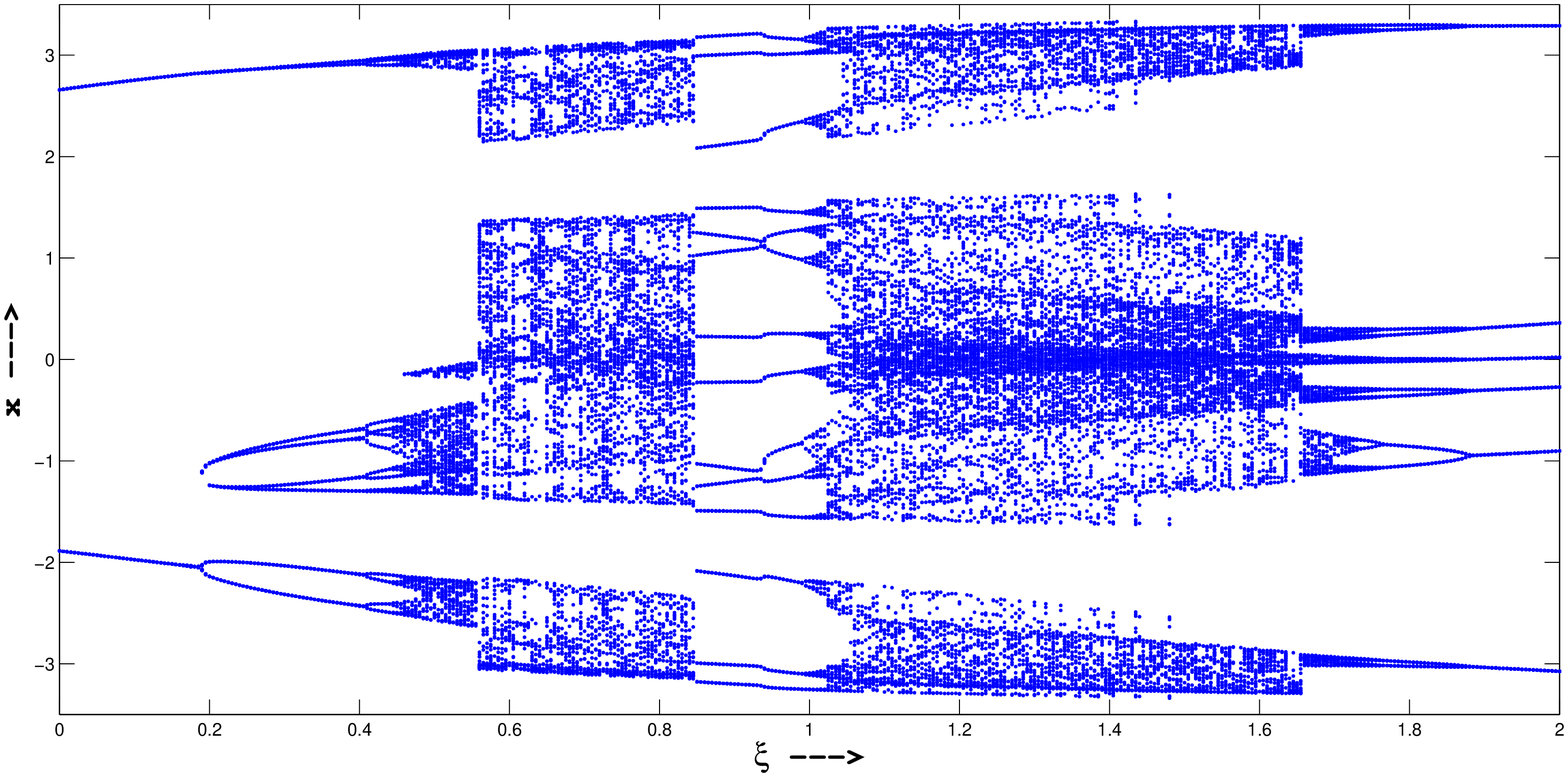}
\caption{ Bifurcation diagram of equation (\ref{ds_1}) with respect to $\xi$ for $f=5.0,~\omega=1.0,~\omega_0^2=0.25,~\alpha=0.2,~\lambda=1.0$}
\end{center}
\end{figure}

\begin{figure}
\begin{center}
\includegraphics[width=1.\textwidth]{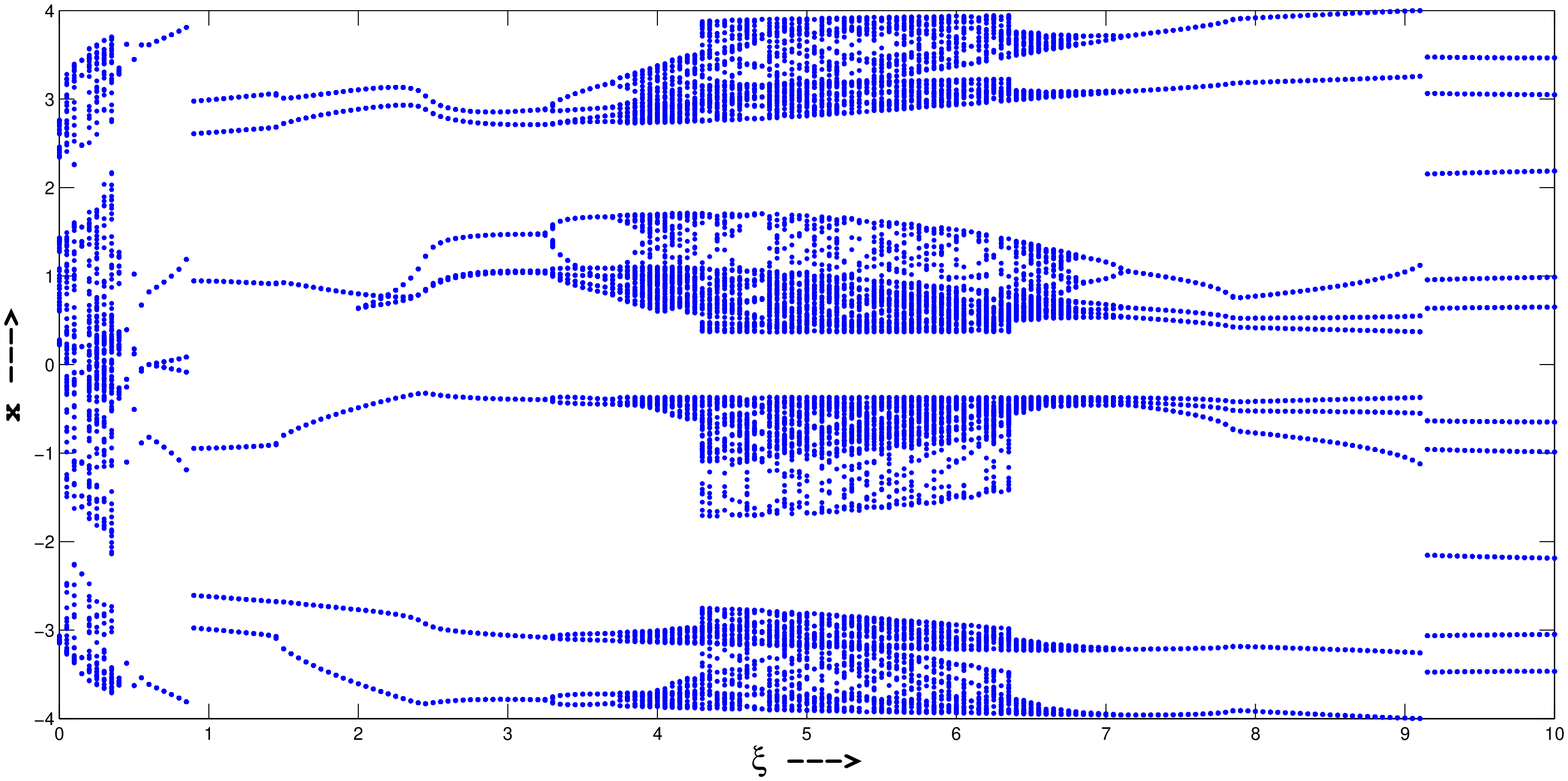}
\caption{ Bifurcation diagram of equation (\ref{ds_1}) with respect to $\xi$ for $ f=8.0,~\omega=1.0,~\omega_0^2=0.25,~\alpha=0.2,~\lambda=1.0$}
\end{center}
\end{figure}

 To summarize, we have demonstrated in this communication that given  any value of $f$, if it is confronted with the PDM parameter $\xi$,  remarkable
 phase transitions occur such as, for instance, from a limit cycle mode to a chaotic regime through a period doubling intermediate phase pointing to the sensitivity of the system (\ref{equation of motion}) due to the interplay between $f$ and $\xi$. The essential distinction between the constant mass and the variable mass case rests in the fact that the presence of the parameter $\xi$ not only enhances the rapidity of such transitions but also initiates complicated nature of bifurcations, of course for a non-zero value of $f$. Conversely, we also encounter transitions such as from a chaotic phase $\rightarrow$
 periodic orbit $\rightarrow$ period doubling $\rightarrow$ chaos $\rightarrow$ regular motion.\\\\
\vspace{0.2cm}
 \textbf{Acknowledgements}\\
 \vspace{0.2cm}
 We thank the anonymous reviewers for their constructive suggestions.


\newpage

  \begin{thebibliography}{99}
 \bibitem{gel} M R Geller and W Kohn \textit{Phys Rev Lett} \textbf{70} (1993) 3103
 \vspace{-0.25cm}
 \bibitem{ser} L Serra and E Lipparini \textit{Europhys Lett} \textbf{40} (1997) 667
 \vspace{-0.25cm}
 \bibitem{rin} P Ring and P Schuck The Nuclear any Body Problem \textit{Springer, NY} 1980
 \vspace{-0.25cm}
 \bibitem{ari} F Arias de Saavedra et al \textit{Phys. Rev.} \textbf{B50} (1994) 4248
 \vspace{-0.25cm}
 \bibitem{pue} A Puente, L Serra and M Casas \textit{Z Phys} \textbf{D31} (1994) 283
 \vspace{-0.25cm}
 \bibitem{bb1} B Bagchi, A Banerjee, C Quesne and V M Tkachuk \textit{J Phys A (Math Gen)} \textbf{38} (2005) 2929
 \vspace{-0.25cm}
 \bibitem{bb2} B Bagchi \textit{J Phys A (Math Theor)} \textbf{40} (2007) F1041
 \vspace{-0.25cm}
 \bibitem{mid} B Midya, B Roy and R Roychoudhury \textit{J Math  Phys} \textbf{51} (2010) 022109
 \vspace{-0.25cm}
 \bibitem{mus} O Mustafa \textit{J Phys A (Math Theor)} \textbf{44} (2011) 355303
 \vspace{-0.25cm}
 \bibitem{cap} M \c{C}apak and B G\"{o}n\"{u}l \textit{J Math Phys} \textbf{52} (2011) 122103
 \vspace{-0.25cm}
 \bibitem{gang} A Ganguly, M V Ioffe and L M Nieto \textit{J Phys A (Math Theor)} \textbf{39} (2006) 14659
 \vspace{-0.25cm}
 \bibitem{ayd} O Aydo\~{g}du, A Arda and R Sever \textit{J Math Phys} \textbf{53} (2011) 042106
 \vspace{-0.25cm}
 \bibitem{chi}  V Chithika Ruby, M Senthilvelan and M Lakshmanan \textit{J Phys A (Math Theor)} \textbf{45} (2012) 382002
 \vspace{-0.25cm}
 \bibitem{car1} J F Carin\~ena, M F Ra\~nada and M Santander \textit{Rep Math Phys} \textbf{54} (2004) 285
 \vspace{-0.25cm}
 \bibitem{car2} J F Carin\~ena, M F Ra\~nada and M Santander \textit{Ann Phys} \textbf{322} (2007) 434
 \vspace{-0.25cm}
 \bibitem{car3} J F Carin\~ena, M F Ra\~nada and M Santander Two important examples of nonlinear oscillators
                \textit{arXiv : math-ph} 0505028
 \vspace{-0.25cm}
 \bibitem{cru1} S Cruz y Cruz, J Negro and L M Nieto \textit{Phys Lett} \textbf{A369} (2007) 400
 \vspace{-0.25cm}
 \bibitem{cru2} S Cruz y Cruz, J Negro and L M Nieto \textit{J Phys : Conf Ser} \textbf{128} (2008) 012053
 \vspace{-0.25cm}
 \bibitem{cru3} S Cruz y Cruz and O Rosas-Ortiz \textit{J Phys A : Math Theor} \textbf{42} (2009) 185205
 \vspace{-0.25cm}
 \bibitem{cru4} S Cruz y Cruz and O Rosas-Ortiz Lagrange equations and Spectrum generators algebras of mechanical sysytems with position-dependent mass\\
                \textit{arXiv :} 1208.2300 
 \vspace{-0.25cm}
 \bibitem{cru5} A Ghose Choudhury and P Guha   SNBNCBS perprint (2012)
 \vspace{-0.25cm}
 \bibitem{sg} S Ghosh and S K Modak \textit{Phys Lett} \textbf{A373} (2009) 1212
 \vspace{-0.25cm}
 \bibitem{lak1} A Venkatesan and M Lakshmanan \textit{Phys Rev} \textbf{E55} (1997) 5134
 \vspace{-0.25cm}
 \bibitem{lak2} P M Mathews and M Lakshmanan \textit{Quart Appl Math} \textit{32} (1974) 215
 \vspace{-0.25cm}
 \bibitem{zno} M Znojil and G Levai Schr$\ddot{\mbox{o}}$dinger equations with indifinite effective mass \textit{arXiv :} 1201.6142
 \vspace{-0.25cm}
 \bibitem{lak3} M Lakshmanan and S Rajasekar Nonlinear dynamics : Integrability, Chaos and Patterns, Advanced Texts in Physics
                \textit{Springer-Verlag,Berlin} (2003)
  \vspace{-0.25cm}
 \bibitem{guc} J Guckenheimer and P Holmes Nonlinear oscillator, Dynamical systems and bifurcation of vector fields
               \textit{Applied Mathematical Science, Springer-Verlag, NY} \textbf{42} (1983)
 \vspace{-0.25cm}
 \bibitem{chu} V Chua Cubic-quintic Duffing oscillators (unpublished)
 \vspace{-0.25cm}
 \bibitem{don} G Donoso and C L Ladera \textit{Eur J Phys} \textbf{33} (2012) 1473


 \end{thebibliography}
 \end{document}